\documentclass[sigconf]{acmart}

\usepackage{adjustbox}
\usepackage{subfig}
\usepackage{array}
\usepackage{graphicx}
\usepackage{clrscode}
\usepackage{balance}
\usepackage{multirow}
\usepackage{float}
\usepackage{color}
\usepackage{amsopn}
\usepackage{mathrsfs}
\usepackage{mathtools}
\usepackage{amsmath}
\usepackage{arydshln}
\usepackage{multicol}
\usepackage{blkarray}
\usepackage{enumerate}
\usepackage{rotating}
\usepackage{booktabs}
\usepackage{algorithm}
\usepackage{algorithmic}

\usepackage{enumitem}

\newcommand{\tabincell}[2]{\begin{tabular}{@{}#1@{}}#2\end{tabular}}

\makeatletter
\g@addto@macro\normalsize{%
  \abovedisplayskip 1pt plus 1pt minus 1pt%
  \belowdisplayskip \abovedisplayskip
  \abovedisplayshortskip 1pt plus1pt  minus1pt%
  \belowdisplayshortskip 1pt plus1pt minus1pt%
}

\makeatother

\renewcommand\footnotetextcopyrightpermission[1]{} 
\title{Value-aware Recommendation based on Reinforced Profit Maximization in E-commerce Systems}

\author{Changhua Pei}
\authornote{Equal contribution. Listing order is random.}
\affiliation{%
  \institution{Alibaba Group}
}
\email{changhua.pch@alibaba-inc.com}

\author{Xinru Yang$^*$}
\authornote{This work was done when Xinru Yang was an intern at Alibaba.}
\affiliation{%
  \institution{Carnegie Mellon University}
}
\email{xyang@andrew.cmu.edu}

\author{Qing Cui}
\affiliation{%
  \institution{Alibaba Group}
}
\email{cuiqing.cq@alibaba-inc.com}

\author{Xiao Lin}
\affiliation{%
  \institution{Alibaba Group}
}
\email{hc.lx@alibaba-inc.com}

\author{Fei Sun}
\affiliation{%
  \institution{Alibaba Group}
}
\email{ofey.sunfei@gmail.com}

\author{Peng Jiang}
\affiliation{%
  \institution{Alibaba Group}
}
\email{jiangpeng.jp@alibaba-inc.com}

\author{Wenwu Ou}
\affiliation{%
  \institution{Alibaba Group}
}
\email{santong.oww@taobao.com}

\author{Yongfeng Zhang}
\authornote{Corresponding author.}
\affiliation{%
  \institution{Rutgers University}
}
\email{zhangyf07@gmail.com}

\begin{document}
%

\begin{abstract}
Existing recommendation algorithms mostly focus on optimizing traditional recommendation measures, such as the accuracy of rating prediction in terms of RMSE or the quality of top-$k$ recommendation lists in terms of precision, recall, MAP, etc. However, an important expectation for commercial recommendation systems is to improve the final revenue/profit of the system. Traditional recommendation targets such as rating prediction and top-$k$ recommendation are not directly related to this goal. 

In this work, we blend the fundamental concepts in online advertising and micro-economics into personalized recommendation for profit maximization. Specifically, we propose value-aware recommendation based on  reinforcement learning, which directly optimizes the economic value of candidate items to generate the recommendation list. In particular, we generalize the basic concept of click conversion rate (CVR) in computational advertising into the conversation rate of an arbitrary user action (XVR) in E-commerce, where the user actions can be clicking, adding to cart, adding to wishlist, etc. In this way, each type of user action is mapped to its monetized economic value. Economic values of different user actions are further integrated as the reward of a ranking list, and reinforcement learning is used to optimize the recommendation list for the maximum total value. Experimental results in both offline benchmarks and online commercial systems verified the improved performance of our framework, in terms of both traditional top-$k$ ranking tasks and the economic profits of the system.
\end{abstract}



\maketitle

\section{Introduction}
Recommender system has become a fundamental service in many online applications. Years of research have witnessed the great advancement on the development of recommendation systems, and various different techniques have been proposed to support better performance in practice, including but not limited to content-based methods  \cite{pazzani2007content}, user/item-based collaborative filtering methods  \cite{ekstrand2011collaborative}, matrix factorization techniques  \cite{koren2009matrix}, and the more recent deep learning-based approaches \cite{zhang2017deep}. Most of the existing  methods focus on two of the most fundamental tasks, namely, rating prediction and top-$k$ recommendation. They usually aim at the optimization of rating or ranking-related measures, such as root mean square error (RMSE), mean average precision (MAP), and many others.

However, an important and sometimes the ultimate goal of commercial recommendation system is to gain revenue and profits for the platform, but traditional recommendation research are not directly related to this goal. They mostly focus on whether or not the algorithm can predict accurate ratings (rating prediction), or if the system can rank the clicked items correctly (top-$k$ recommendation). Nevertheless, researchers have shown that improved performance on rating prediction does not necessarily improve the top-$k$ recommendation performance in terms of user purchase  \cite{cremonesi2010performance}. And even better performance on top-$k$ recommendation does not necessarily improve the profit of the system  \cite{zhang2016economic}, because the purchased recommendations may not convert into the best profit due to their difference in price. For example, by recommending daily necessities, the system has a better chance of getting the user to purchase the recommendation, but the expected profit achieved from the purchase may be smaller than if a luxury good was purchased, though the latter has a smaller chance of being purchased.

In this work, we propose value-aware recommendation for profit maximization, which directly optimizes the expected profit of a recommendation list. The basic idea is to balance the unit profit of a recommendation and the chance that the recommendation is purchased, so that we can find the recommendation list of the maximum expected profit. To achieve this goal, we generalize the key concept of click conversion rates (CVR)  \cite{lee2012estimating,agarwal2011location} in online advertising to the conversion rate of arbitrary user actions (XVR), which scales each user action into the monetized profit of the system based on large-scale user behavior logs.

Once the user actions are converted into expected profits, we further design an aggregated reward function based on the monetized user actions, and then adopt reinforcement learning to optimize the recommendation list towards the maximized expected profit for the system. For training, we developed a benchmark dataset by collecting the user behaviors in a real-world E-commerce system. Furthermore, we conducted online experiments on this E-commerce system to validate the online performance of value-aware recommendation in terms of both ranking performance and system profits.

The key contributions of the paper can be summarized as follows:
\begin{itemize}[leftmargin=*,topsep=0pt,noitemsep]
\item We propose \textit{value-aware recommendation} to maximize the profit of a recommendation system directly, which to the best of our knowledge, is the first time to explicitly optimize the profit of the personalized recommendation list in large-scale online system.
\item We propose to generalize the basic concept of click conversation rate into the conversion rate of arbitrary user actions, so that each user action can be monetized into the profit of the system.
\item By aggregating the monetized user actions into a unified profit reward, we develop a reinforcement learning-based algorithm for value-aware recommendation.
\item We conduct both offline experiments with benchmark datasets and online experiments with real-world commercial users to validate the performance of value-aware recommendation on ranking and profit maximization.
\end{itemize}

\section{Related Work}\label{sec:related}
In this section, we introduce the key related work in terms of three perspectives: economic recommendation, computational advertising, and reinforcement learning.

\subsection{Economic Recommendation}
For a long time, recommendation system research has been working on rating- or ranking-related tasks such as rating prediction, top-$k$ recommendation, sequential recommendation, etc., and a lot of successful models have been proposed, which greatly advanced the performance of recommendation systems. To name a few, this includes content-based methods \cite{pazzani2007content}, user/item-based collaborative filtering \cite{ekstrand2011collaborative}, shallow latent factor models such as matrix factorization  \cite{koren2009matrix,lee2001algorithms,mnih2008probabilistic,rendle2014improving}, and more recent deep learning based methods  \cite{zhang2017deep,wang2015collaborative,he2017neural}. However, the related methods seldom consider the economic value/profit that a recommendation list brings about to the system, although this is one of the most important goals for real-world commercial recommender systems. Some recent research on economic recommendation has begun to take care of the economic value of recommendation\cite{vams,value-aware1,value-aware2},. For example,  \cite{zhang2016economic} proposed to maximize social surplus for recommendation, and \cite{zhao2017multi} proposed to learn the substitutional and complementary relations between items for utility maximization. However, none of the above methods explicitly maximizes the expected profits of a recommendation list to generate recommendations. 

\subsection{Computational Advertising}

Currently, computational advertising~\cite{dave2014computational} has been playing a crucial role in online shopping platforms, rendering it essential to apply an intelligent advertising strategy that can maximize the profits by ranking numerous advertisements in the best order \cite{ghose2009empirical,broder2007semantic,goldfarb2011online}. In this scenario, the goal of advertisers is to maximize clicks and conversions such as item purchase. Therefore, cost-per-click (CPC) and cost-per-action (CPA) models are often adopted to optimize the click-through-rate (CTR) and the conversion-rate (CVR), respectively. Although technically there are a lot of analogy between advertising and recommendation, there is not much work trying to bridge these two principles. This work is one of our first attempts to integrate the economic consideration in adversting into recommender systems for value-aware recommendation.

\subsection{Reinforcement Learning}
Reinforcement Learning (RL) aims at automatically learning an optimal strategy from the sequential interactions between the agent and the environment by trial-and-error~\cite{RL}. 
Some existed works have made efforts on exploiting reinforcement learning for recommendation systems. For instance, conventional RL methods such as POMDP (Partially Observable Markov Decision Processes) \cite{yxr25} and Q-learning \cite{yxr27} estimate the transition probability and store the Q-value table in modeling. However, they soon become inefficient with the sharp increase in the number of items for recommendation. Meanwhile, other approaches have been proven useful with an effective utilization of deep reinforcement learning. For instance, \cite{listwise} employs Actor-Critic framework to learn the optimal strategy by a online simulator. Furthermore, \cite{negative} considers both positive and negative feedback from users recent behaviors to help find optimal strategy. \citeN{feng2018learning} uses the multi-agent reinforcement learning to optimize the multi-scenario ranking. Meanwhile, \cite{pagewise} adopts RL to recommend items on a 2-D page instead of showing one single item each time.

To mitigate the performance degradation due to high-variance and biased estimation of the reward, \cite{kddali} provides a stratified random sampling and an approximate regretted reward to enhance the robustness of the model. Similarly, \cite{ecommerce} introduces DPG-FBE algorithm to maintain an approximate model of the environment to perform reliable updates of value functions.

\section{Generalized Conversion Rate}
\label{sec:xvr}
In order to maximize the profit of recommender system, we need to measure the the value of different user actions on the platform. In this section, we introduce how to generalize the basic concept of click conversion rate to rescale each user action into monetized profit. Here we use Gross Merchandise Volume (GMV) to measure the profit of the system. GMV is a measure used in online retailing, which indicates the total sales dollar value of the merchandise sold through a particular marketplace over a certain time frame\footnote{https://en.wikipedia.org/wiki/Gross\_merchandise\_volume}. We use GMV and profit interchangeably aftermentioned.


\subsection{Conversion Rate of Clicks}
Click conversion rate (CVR) is widely used in online advertising to estimate the potential profit of an advertising item to the advertising platform \cite{cvr-ad,cvr-ali}. It is usually defined as the result of dividing the number of conversions by the number of total clicks that can be tracked to a conversion during the same period of time. In E-commerce system, it is difficult to learn an optimal recommendation strategy by simply calculating the profits of transactions because purchasing an item is a relatively sparse action of users. Instead, users tend to click an item much more frequently. CVR provides a way to map the dense action of clicking into a scaled range of profits. For E-commerce system, CVR is the first step towards value-aware recommendation\cite{jiang2015life,xia2018modeling}. Eq,~\ref{eq:backgroud} shows the total estimated GMV brought by click, which shows that CVR plays a crucial role in profit maximization, 
\begin{equation}
    E[GMV]=\sum_{i}I(click, i) \cdot \mathit{CVR}(i) \cdot price(i)
    \label{eq:backgroud}
\end{equation}
where price($i$) is the unit price of item $i$, $I(click,i)$ is an indicator of the occurrence of click on item $i$. If item $i$ is clicked, $I(click,i)=1$, otherwise, $0$.

\subsection{Generalization to other Actions}
\begin{figure}[t]
    \centering
    \includegraphics[scale=0.4]{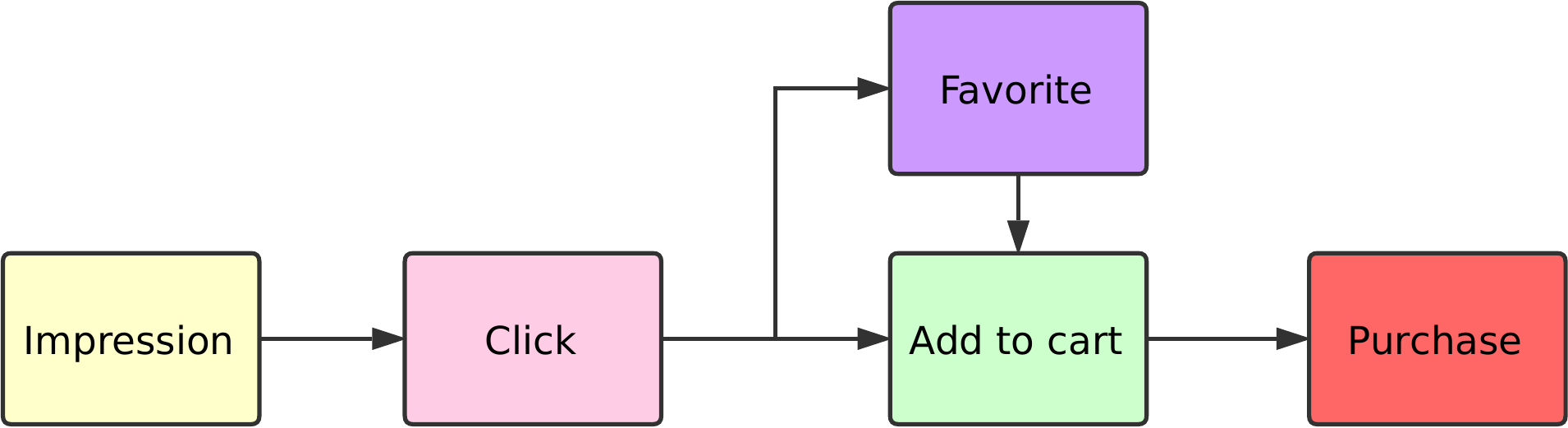}
    \caption{A flow of the user behavior series}
    \label{fig:state_tran}
\end{figure}

In E-commerce platform, a series of behaviors of a user is often formed step by step with transforming probabilities. In Figure ~\ref{fig:state_tran}, the general shopping process is simplified into four steps. First, the user sees something of interests and gets an impression on the item, then the user will click to see more details and to eventually add it to the cart. The user can also add it to the wishlist so he can add it to the cart later. Eventually, some items in the cart will be purchased together and contribute to final profits. Adding to wishlist and adding to cart reveal a strong desire for the item even if no transaction is made eventually. Therefore, these actions are also considered to contribute to the value.



Here we generalize the concept of CVR to XVR to map each type of user action $x$ to its monetized economic value. XVR is defined as the possibility of transition from certain action $x$ to purchase. Using XVR, we can estimate the total profits of different actions on platform. By generalizing Eq.~\ref{eq:backgroud}, we get Eq.~\ref{eq:gmv}, 
\begin{equation}
    E[GMV] = \sum_{x,i} V(x,i) = \sum_{x,i} I(x, i) \cdot \mathit{XVR}(i) \cdot price(i)
    \label{eq:gmv}
\end{equation}
where E[GMV] denotes the expected GMV of the platform, $x$ represents certain action including \textit{click}, \textit{add to cart}, \textit{add to wishlist}, and \textit{purchase}. Here $V(x,i){=}I(x,i) \cdot \mathit{XVR}(i) \cdot price(i)$ is the expected profit given user has an action $x$.

\begin{figure*}[hbtp]
    \centering
    \includegraphics[scale=0.5]{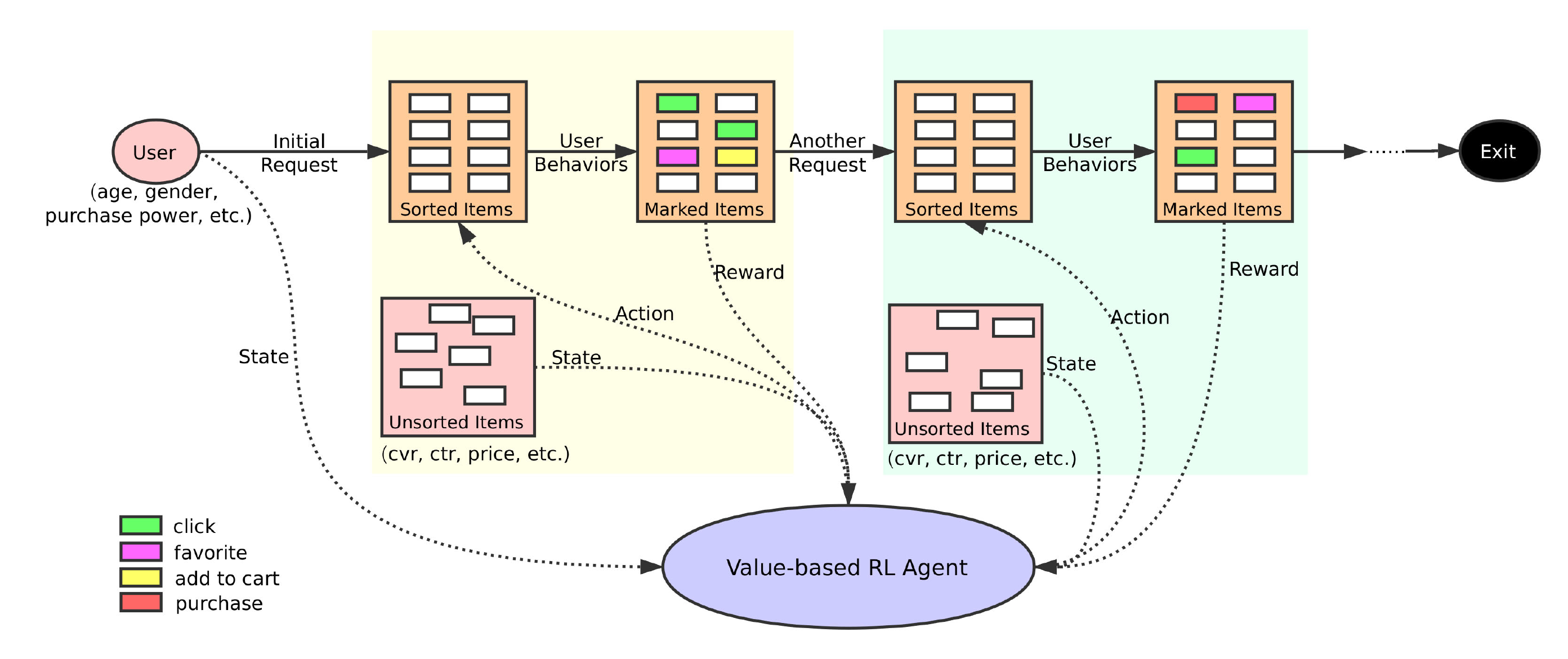}
    \caption{Structure of the model. When user initiates a request, the system will generate a recommendation list, and the user will interact with the list with certain actions. These Actions will be used to calculate the reward to update the RL Agent.}
    \label{fig:es}
\end{figure*}

\section{Reinforced Profit Optimization}
\label{sec:algorithm}

The main goal of this work is to maximize the total profit of recommender system in an E-commerce scenario. Here the profit indicate a total sales dollar value for merchandise sold online in a certain time frame.  
Using the generalized conversation rate, we can transfer all user actions to monetized profit and maximize it. Reinforcement learning (RL) aims to learn a policy that maximizes the accumulated reward. The profit can naturally be used as the reward with which we can find better actions (ranking policy). Besides, the view of interact between the environment (users) and RL agent fits our recommendation scenario well. 
Therefore, in this paper, we choose reinforcement learning, evolution strategy more specifically\cite{es},  as our algorithm for its simplicity and effectiveness.

\subsection{Reinforcement Learning Modeling}

Figure~\ref{fig:es} illustrates how our value-based RL agent interacts with the user to maximize the profit. First, the user comes to the system at some time and makes an initial request, the system receives some information about the user as well as the candidate items as a state and responds to the request, which virtually means taking an action by generating an order to rank items. Next, the user shows a certain behaviors regarding the items such as click, add to wishlist, add to cart or purchase, which is used to calculate the reward to the system for what kind of action it takes.

In our recommendation platform, items are shown in cascade on a mobile App one by one. Each time the user initiates a request, 50 items are recommended to him/her. As user scrolls down the list and have seen all 50 items, a new request is triggered. This process is repeated until the user leaves the App or return the top of the cascade, labelled as "exit" in Figure~\ref{fig:es}. We use a metric called "pageid" to distinguish different requests in this interaction, similar to the concept of "page" to a search engine. As the user and the system interact with each other, the system learns how to respond to the state to obtain an optimized accumulative reward. 


\begin{figure*}
     \centering
     \subfloat[][Average GMV.]{\includegraphics[width=.35\linewidth]{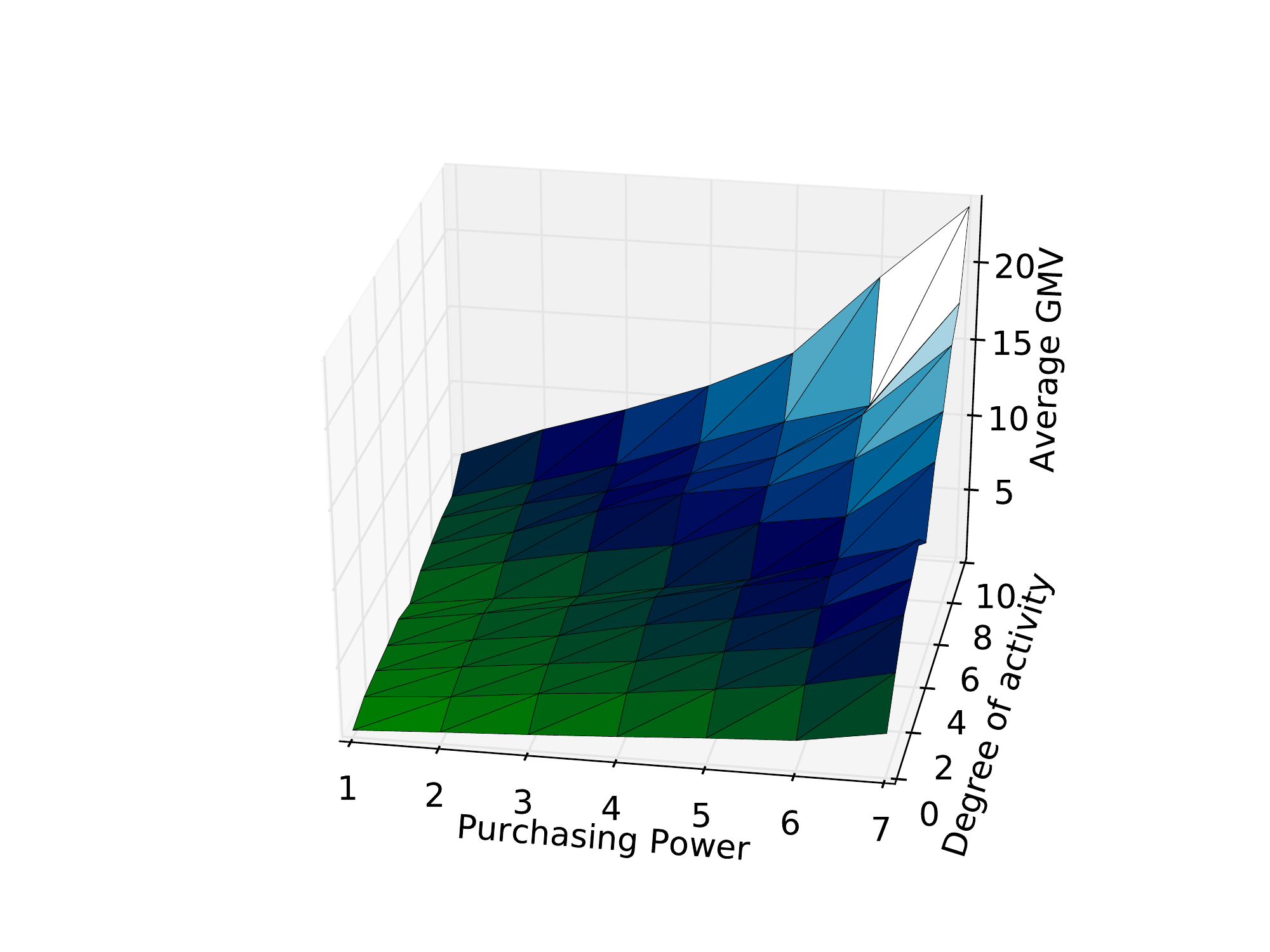}\label{fig:side:a}}
     \subfloat[][Number of Users.]{\includegraphics[width=.35\linewidth]{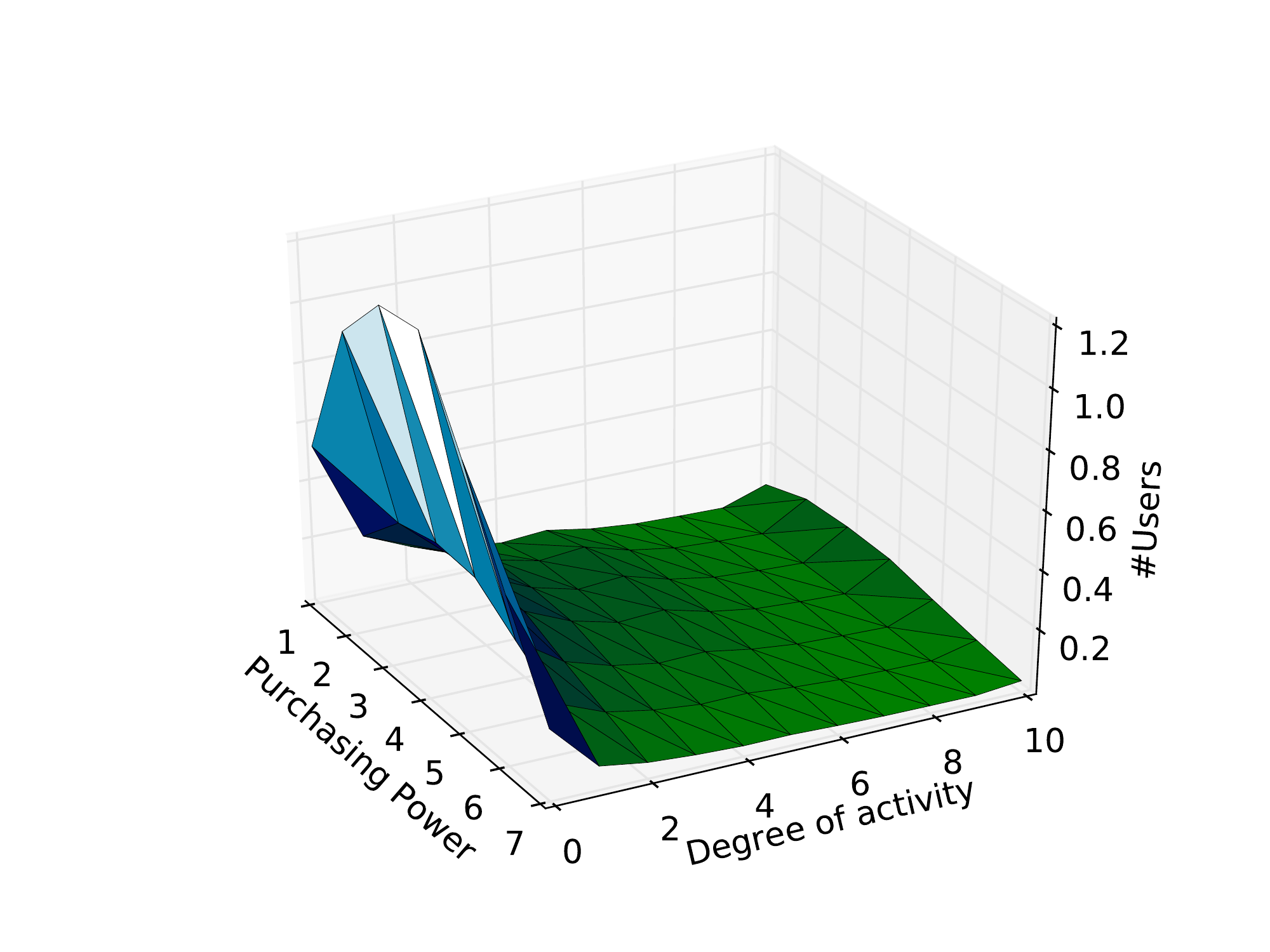}\label{fig:side:b}}
     \subfloat[][Summed GMV.]{\includegraphics[width=.35\linewidth]{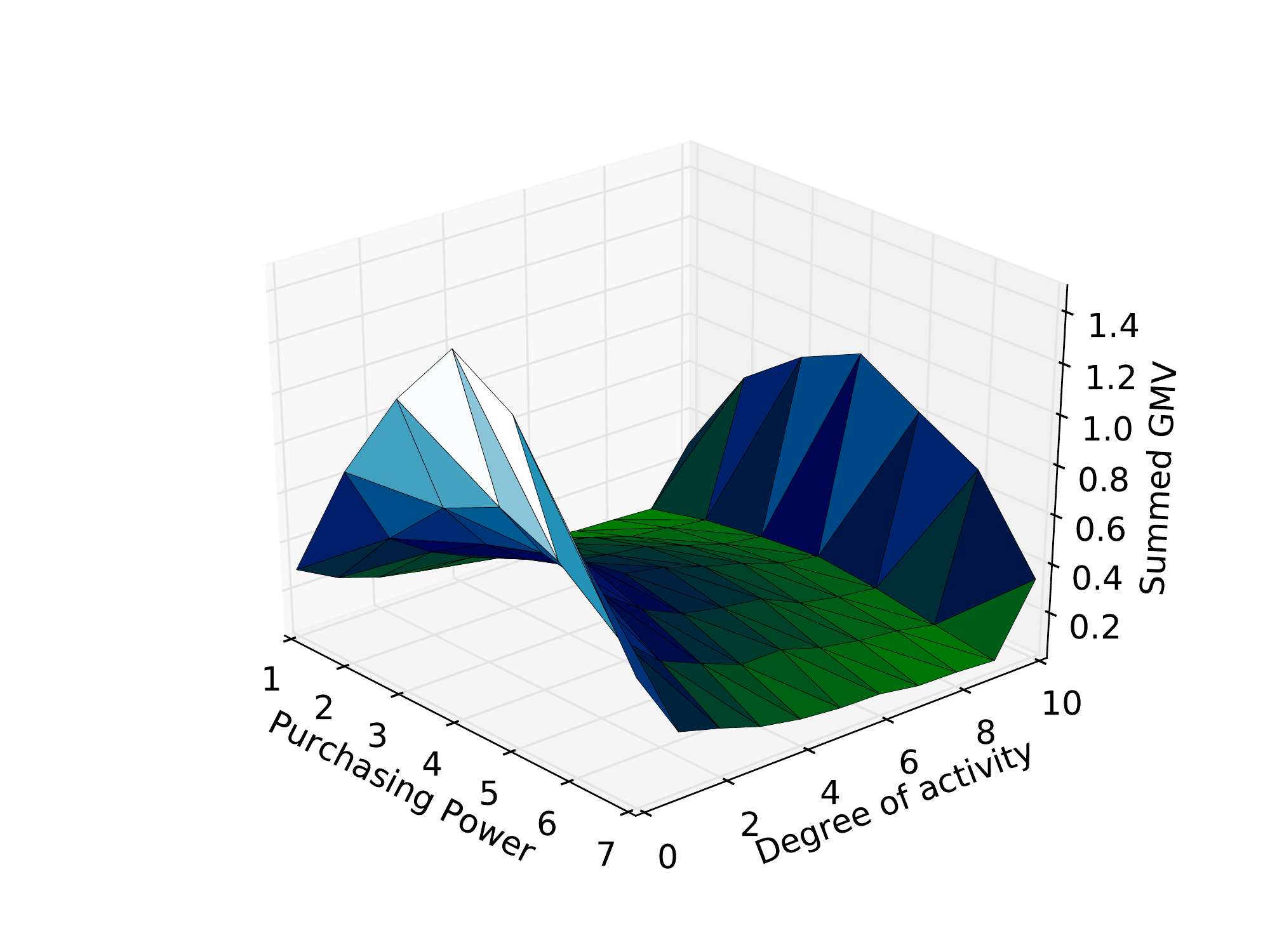}\label{fig:side:c}}
     \caption{Average/Summed GMV and the number of users under different level of purchasing power and degree of activity.}
     \label{fig:side}
\end{figure*}


\textbf{States:}~
States are used to describe an user's request, including user-side, item-side, and contextual features. The user-side features include \textit{age}, \textit{gender}, \textit{purchase power}. The item-side features are \textit{ctr}, \textit{cvr}, \textit{price}. The contextual features include \textit{pageid}, \textit{request time}. In total, the states in our model can be formulated as <\textit{age}, \textit{gender}, \textit{purchase power}, $ctr_0$, ..., $ctr_{49}$, $cvr_0$, ..., $cvr_{49}$, $price_0$, ..., $price_{49}$, \textit{pageid}, \textit{request time}>, the subscript $i$ denotes the corresponding features of $i^{th}$ item in the corresponding page. 

All these features work together to influence the final GMV. For example, purchase power is calculated by how many and how expensive the bought items are. The higher the level of purchase power is, the more money is spent on the platform. Fig.~\ref{fig:side:a}, \ref{fig:side:b}, and \ref{fig:side:c} show average GMV per user, number of users and summed GMV under different purchasing power and degree of activity. Degree of activity indicates the time a user spends on the platform. Fig.~\ref{fig:side:a} shows users with higher purchase power have higher average GMV. Thus the model tends to recommend items with higher price to user who has high purchase power. However, for summed GMV, users with middle level of purchase power contribute most because the large number of users, as shown in Fig.~\ref{fig:side:b}. 

\begin{figure}[t!]
    \centering
    \includegraphics[scale=0.35]{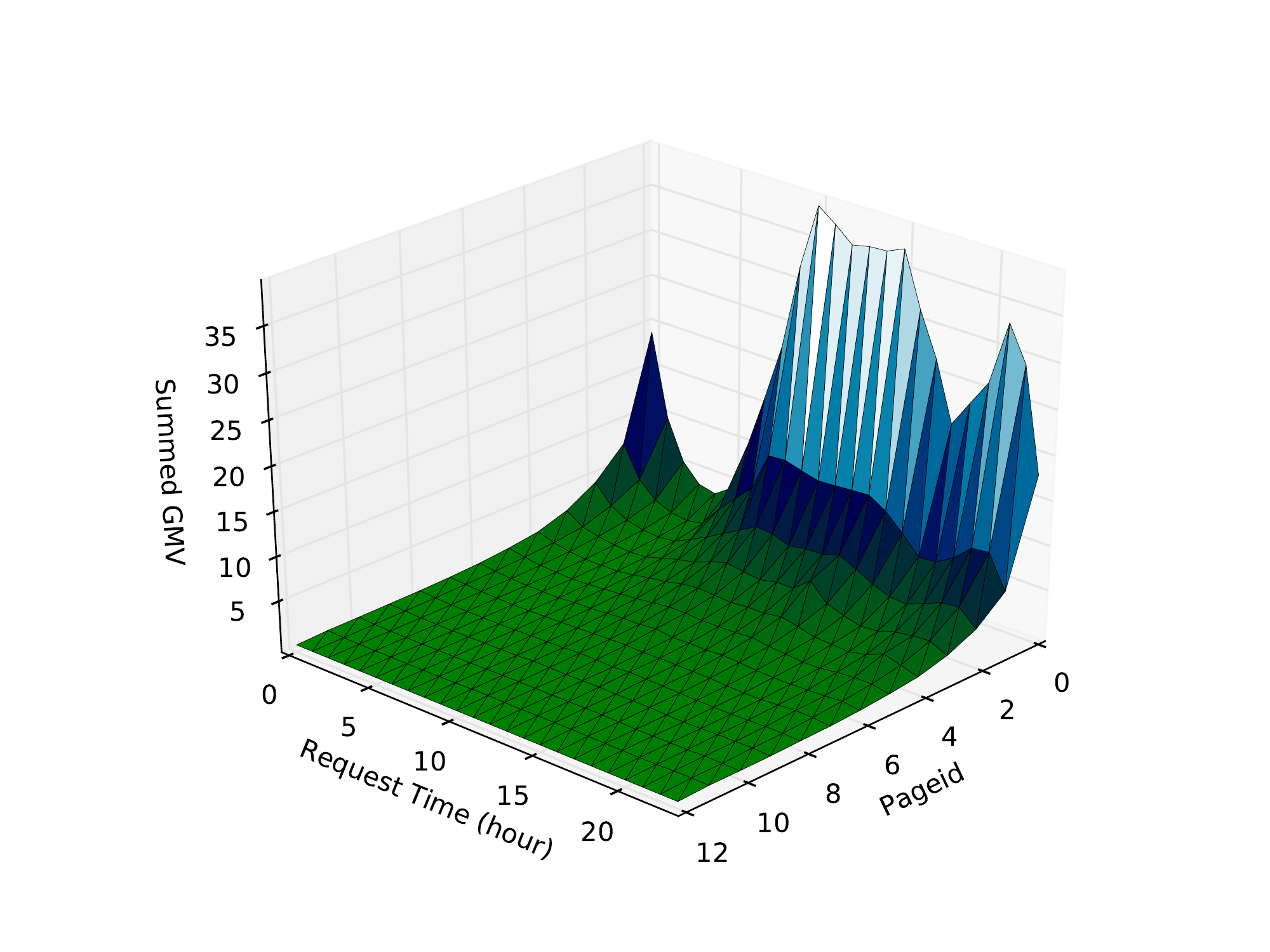}
    \caption{GMV across different pages and time.}
    \label{fig:pageid_hh}
\end{figure}

Request time (labelled as hour in Fig.~\ref{fig:pageid_hh}) and pageid are also features used in our model. Fig.~\ref{fig:pageid_hh} shows that the the first page whose pageid equal to 0 contributes most to the summed GMV, as user scrolls downwards, pages with larger pageid have smaller possibilities to have impression on users. Thus the model should recommend items with higher conversion rate in the first page. Besides, model will learn different weights at different hour as Fig.~\ref{fig:pageid_hh} show the summed GMV varies with request time.

\textbf{Actions:}~
Action is the coefficient vector in the ranking formular which decides the order of candidate items. We design a ranking formula shown in Eq.~\ref{eq:rankscore} to order the items,
\begin{equation}
	rankscore(i) {=}\sum_{x \in \mathcal{A}} P(x,i) ^ {\alpha_x} {\cdot} \mathit{XVR(i)} ^ {\beta_x } {\cdot} price(i) ^ \gamma
	\label{eq:rankscore}
\end{equation}
where $\mathcal{A}$ is the user action set which contains click, add to cart, add to wishlist. $P(x,i)$ is the possibility for a given user has action $x$ on item $i$. $XVR(i)$ is the generalized conversion rate of item $i$. In total, the action can be formulated as ${<}\alpha_{\mathit{click}},\alpha_{cart},\alpha_{\mathit{fav.}},\beta_{\mathit{click}},\beta_{\mathit{cart}},\beta_{\mathit{fav.}},\gamma{>}$.

If we only consider click in the action set $\mathcal{A}$, rankscore can be simplified to Eq.~\ref{eq:rankscore_click}. The corresponding \textit{action} becomes ${<}\alpha_{\mathit{click}},\beta_{\mathit{click}},\gamma{>}$.
\begin{equation}
\begin{split}
rankscore(i) &{=} P(click,i) ^ {\alpha_{click}} {\cdot} \mathit{XVR} ^ {\beta_{click}} {\cdot} price(i) ^ \gamma \\
&{=} ctr(i) ^ {\alpha_{click}}  {\cdot} cvr(i) ^ {\beta_{click}} {\cdot} price(i) ^ \gamma
\end{split}
	\label{eq:rankscore_click}
\end{equation}
Changing an action means to change the coefficient vector for the ranking formula.

\textbf{Reward:}~
In our value-based method, we use the expected profit as reward, which contains the monetized profit converted from all kinds of user actions. Based the definition of the expected GMV in Eq.~\ref{eq:gmv}, for a given item $i$, the reward $R_i$ can be defined as:
\begin{equation}
	R_{i} = V(click,i) + V(fav.,i) +V(cart,i) + V(pay,i)
	\label{eq:value}
\end{equation}

In this paper, unless specifically mentioned, we mainly use click and pay in reward, \textit{i.e.}, $R_i {=} V(click,i) {+} V(pay,i)$. In the last section, we evaluate the performance when taking adding to cart and adding to wishlist into reward.
The total reward of a recommendation list in a page that contains T items can be defined as $R_{page} = \sum_{i=0}^{T}R_{i}$.





\subsection{Model Training}
\subsubsection{Offline Reward}
In the above section, the reward of the model $R_{page}$ is directly calculated with user's feedback online. This requires that the model is deployed online directly and learns the policy from the real-time data streaming. However, the online traffic is expensive. It is risky to let RL model learns directly online before we validate the effectiveness of the value-based method. To overcome this problem, we propose a simulated environment which leverage the historical data to approximate the feedback of the users. The simulated environment adopts a simple idea of NDCG (normalized discounted cumulative gain) that good policy should rank the item user actually clicked and paid in the front. In this way, an offline reward $R^{'}_{page}$ is used to evaluate actions as following: 
\begin{equation}
R^{'}_{page} = \sum_{i=0}^{T} R_{i} * W_{\pi(i)}
\label{eq:ndcg}
\end{equation}
For a given page with $T$ items $1,\cdots,T$, the ranking policy can generate an order list, where $\pi(i)=k$ means item $i$ is ranked at position $k$. We assign a discounted weight for each position $k$ and higher position has greater weight, i.e. $W_{\pi(i)}$. Then we represent the reward as the weighted sum of item reward. In this paper, $W_{\pi(i)}$ is represented by exponential function which is $W_{\pi(i)}=\exp(\pi(i))$.

\subsubsection{Learning Algorithm}
As the main goal of this paper is to evaluate the effectiveness of the value-aware recommendation, we choose the simplest model in reinforcement learning model family to exclude the influence of complex models. Finally we adopt evolution strategy algorithm (abbreviated ES algorithm). We train the model offline and deploy it online to perform A/B tests. The following introduces how the ES algorithm learns in the training process.

Our task is to learn a policy which maps state to action that maximizes the expected reward. To keep it simple, we use a linear model as the policy and $\theta$ is the parameters to be optimized. The ES algorithm starts from initial parameter $\theta_0$ and then updates $\theta_t$ in the loop where $t{=}1,2,\dots$. At each iteration loop $t$, the user's request for a new page and pass the state vector $s_t$ to the model. The algorithm generates $n$ baby-agent and each baby-agent $j$ samples a set of increments $\epsilon_t^j$ from a normal distribution. Each model variant with parameter $\theta_t + \epsilon_t^j$ maps the state to an action. Each action orders the list in a different way and the user reacts to different ranking polices by clicking or purchasing certain items in the list. We calculate the reward of each policy $j$ using Eq.~\ref{eq:ndcg} and labeled as $R^{'}_{page}(j) = F(\theta_t + \epsilon_t^j)$. We collect all the rewards together and update $\theta_t$ eventually:
\begin{equation}
\theta_{t+1} \xleftarrow{} \theta_{t}+\alpha_{lr}\frac{1}{n\sigma}\sum_{j=1}^{n}R^{'}_{page}(j)
\label{eq:updates}
\end{equation}
where $\sigma$ and $\alpha_{lr}$ are hyper parameters, $\sigma$ is the noise standard deviation and $\alpha_{lr}$ is the learning rate. 

\section{Experiments}
\label{sec:experiment}

In this section, we first introduce the settings of our benchmark dataset used for training or offline evaluation. Then we evaluate our method both online and offline together with baselines. \footnote{The code and dataset of the paper are released at https://github.com/rec-agent/rec-rl.}

\subsection{Benchmark Dataset}
Our benchmark dataset is collected from real-world E-commerce platform and can simulate the interactive environment and users' behaviors. Each piece of data is comprised of two parts: the \textit{states} and users' feedback: click, add to cart, add to wishlist, purchase. Table~\ref{tab:dataset} lists the details of the dataset.

\begin{table}[t]
\centering
\caption{Overview of the benchmark Dataset}
\begin{tabular}{lr}
\toprule
 clause & size(x$10^6$) \\
\midrule
\#distinct users & 49  \\
\#distinct items & 200 \\
\#requests & 500 \\
\#clicks & 670 \\
\#adding to carts & 60  \\
\#adding to wishlists & 30  \\
\#purchases & 3 \\
\bottomrule
\end{tabular}
\label{tab:dataset}
\end{table}


\subsection{Experimental Setup}

\begin{table*}[htb]
\centering
\caption{Offline Evaluation Results (p-value < 0.005). Numbers in the brackets are improvements compared to Item-based CF and LR-based LTR.}
\begin{tabular}{lrrrrr}
\toprule
 & E[GMV] & Average $R^{'}_{page}$ & Precision$@$20(\%) & Recall$@$20(\%) & MAP$@$20(\%) \\
\midrule
Item-based CF  & 0.40 & 19.73 & 2.96  &  27.93 & 8.69 \\
LR-based LTR   & 0.49 (22.5\%/-) & 25.87 (31.1\%/-) & 3.45 (16.6\%/-)  &  32.42 (16.1\%/-) & 11.04 (27.0\%/-)  \\
DNN-based LTR  & 0.50 (25.0\%/2.0\%) & 25.88 (31.2\%/0.04\%) & 3.65 (23.3\%/5.8\%) &  34.01 (21.8\%/4.9\%) & 12.27 (41.2\%/11.1\%)   \\
\textbf{Value-based RL} & 0.53 (32.5\%/8.2\%) & 27.78 (40.8\%/7.4\%) & 3.74 (26.4\%/8.4\%) & 34.82 (24.7\%/7.4\%) & 12.36 (42.2\%/12.0\%)\\
\bottomrule
\end{tabular}
\label{tab:offline}
\end{table*}

\begin{table}[htb]
\vspace{7pt}
\caption{Online Evaluation Results (p-value < 0.005).}
\centering
\begin{adjustbox}{width=\linewidth}
\begin{tabular}{lrrr}
\toprule
 & GMV & CTR(\%) & IPV \\ 
\midrule
Item-based CF              & 7.57  & 3.04  & 2.48  \\
LR-based LTR              & 8.95 (18.3\%/-) & 3.26 (7.4\%/-) & 2.67 (7.5\%/-) \\
DNN-based LTR             &  9.06 (19.7\%/1.2\%) & 3.28 (8.1\%/0.6\%) & 2.69 (8.3\%/0.7\%)  \\
\textbf{Value-based RL}  & 9.68 (27.9\%/8.2\%) & 3.29
(8.2\%/0.9\%) & 2.70 (8.8\%/1.1\%)  \\ 
\bottomrule
\end{tabular}
\end{adjustbox}
\label{tab:online}
\vspace{5pt}
\end{table}

\subsubsection{\textbf{Baselines}:}
We use the following three methods as baselines to compare with our value-based RL method: Item-based Collaborative Filtering,  LR-based learning to rank(LTR), DNN-based learning to rank.

\begin{figure}[t]
    \centering
    \includegraphics[scale=0.35]{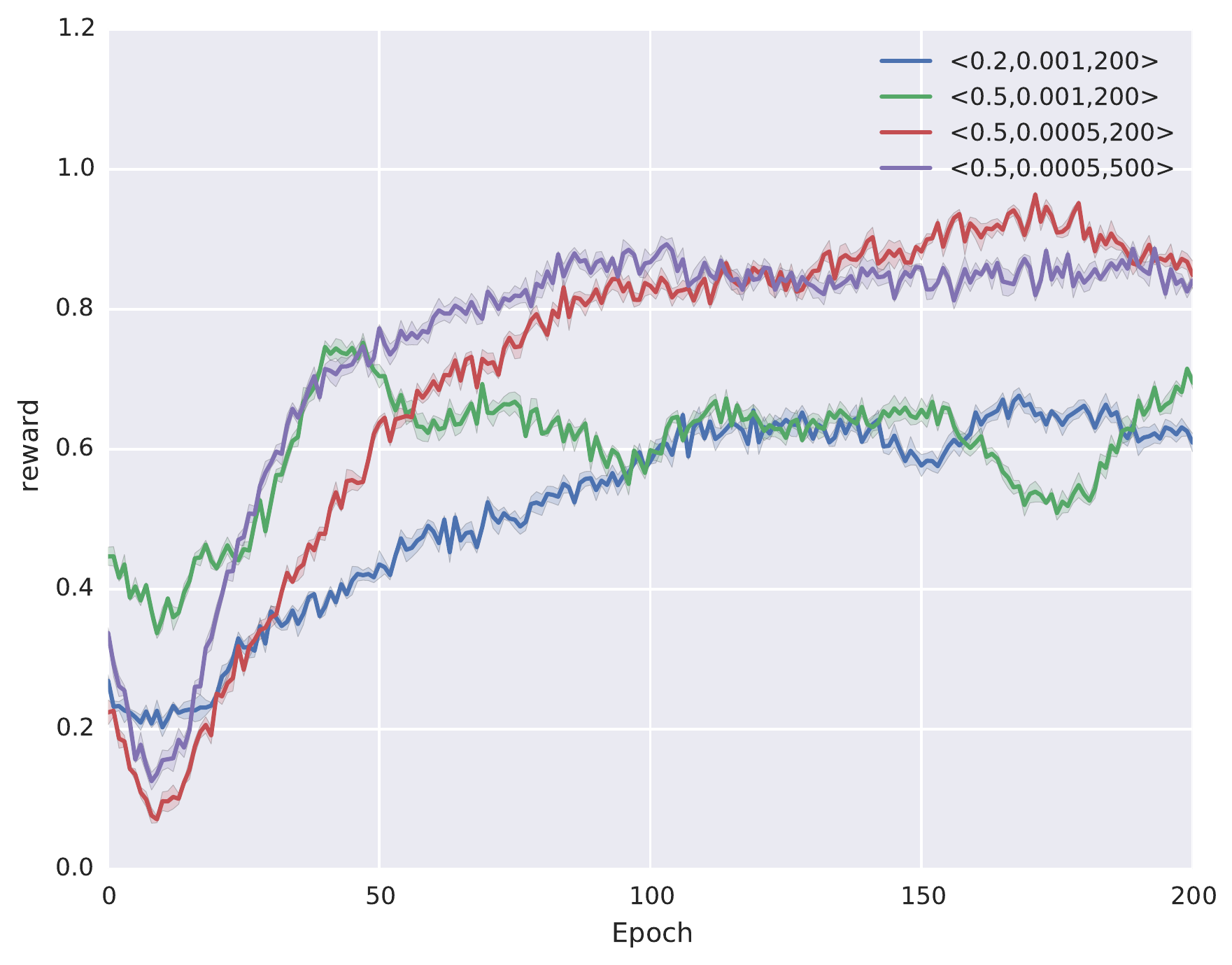}
    \caption{The training curve under different parameters: $<$noise  standard deviation, learning rate, batch size$>$}
    \label{fig:reward_aio}
\end{figure}

\begin{itemize}[leftmargin=*,topsep=0pt]
    \item \textbf{Item-based CF}: The classic Item-based CF~\cite{sarwar2001item} is widely used for recommendation. compared to user-based CF \cite{herlocker2017algorithmic}, Item-based CF is more scalable and accurate in E-commerce recommendation because the relationships between items are more stable. In this paper, we use a fine-tuned item-based collaborative filtering method as one of our baselines. Other CF methods such as matrix factorization can also be used as baselines. However, as they are all non value-aware methods, we omit them in this paper for simplicity.
    \item \textbf{LR-based LTR}: The Point-wise Learning-To-Rank paradigm \cite{Liu:2009:LRI:1618303.1618304} is adopted to learn the rankings of recommended items with Logistic Regression(LR) as the ranking model. We use clicked and purchased items as positive samples to learn a binary classification model, and the price is used as the weight of loss. To reduce the variance of price and make the optimization stable, we actually use $\log(price)$ as the weight. For fair comparison, here we employ the same feature set as our proposed value-based RL.
    There are three hyper-parameters involved in the training of LR-based LTR, \textit{i.e.}, the $\ell_2$ regularization weight, the learning rate and the batch size. These parameters are set to 0, 0.5 and 128 after tuning. 
    We also test a repression model for predicting the profit.
    However, it cannot beat classification models like LR.
    This is consistent with previous work~\cite{gunawardana2009survey}, regression model may not be suitable for ranking problems in recommender system.
    
    \item \textbf{DNN-based LTR}: This method is the same as LR-based LTR, except that the ranking model is a Deep Neural Network (DNN) instead. The neural network involves two hidden layers with 32 and 16 neurons respectively. The $\ell_2$ regularization weight, the learning rate and the batch size are set to 0, 0.05 and 1024. 
    We also tested network structures including RNN, attention networks, and Wide\&Deep network. Optimizer, activation functions, normalization, and other hyper parameters are tuned by greedy search. However, the improvement of these networks is not significant on very large dataset. 
    
\end{itemize}

\subsubsection{\textbf{Value-based RL}:}
Different hyper parameters are tried by a greedy search to find the better training performance. In ES algorithm, under the same training set, the larger reward a model can achieve, the better the model is. We name some important parameters in Fig.~\ref{fig:reward_aio} to illustrate this. Having a larger deviation (from 0.2 to 0.5) can greatly improve the maximum reward the model can achieve. However, the model is not stable so that the reward will drop after some iterations. We adjust the  learning rate in Adam optimizer from 0.001 to 0.0005 and the training curve become stable. We also try larger batch(from 200 to 500), the improvement is not that much. Finally, $<$0.5,0.0005,200$>$ are used as noise standard deviation, learning rate and batch size in our RL model respectively.

\vspace{3pt}
\subsection{Key Results}
We evaluate our method with both offline and online metrics, see Table~\ref{tab:offline} and Table~\ref{tab:online}\footnote{Normalized related value instead of the original absolute value of the online measures are shown for business reason. The relative improvement is consistent with real experimental results.}. 

\vspace{5pt}
\textbf{Offline Evaluation:} In offline evaluation, we train the model on the data of one week and evaluate it on the data of the following week. Precision, recall and MAP within the top 20 items is used to measure the performance. Besides the traditional metrics, average $R^{'}_{page}$ (Eq.~\ref{eq:ndcg}) and E[GMV] (Eq.~\ref{eq:gmv}) are used as value-based metrics.

Table~\ref{tab:offline} shows that the Learning-To-Rank (LTR) baseline outperforms item-based CF in all measures which indicates the LTR methods can predict the profit more effectively. We leverage price in two ways. On one hand, LTR methods use price as one of the features. On the other hand, LTR methods use price as the weight for loss in the learning process. Our value-based approach has 6.0\% and 7.3\% improvements in E[GMV] and $R^{'}_{page}$ respectively when compared to DNN-based LTR. Besides, Value-based RL achieves better performance in precision(2.5\%), recall(2.4\%) and MAP(0.7\%) than DNN-based LTR with same features. Our approach can precisely monetize an arbitrary user action into the profit by generalizing the basic concept of click conversion rate (CVR) to CXR. 
 

\vspace{5pt}
\textbf{Online Evaluation:} An online A/B test is used to evaluate different methods. Metrics we use in A/B test are ctr, IPV and GMV. IPV is the absolute number of clicked items on the platform, which acts a supplement metric to ctr. To make the online evaluation comparable, each bucket in A/B test has the same number of users and contains over millions of users.

Table~\ref{tab:online} shows the average results for one week. Online evaluation shows results are consistent with the offline evaluation. The DNN-based LTR  improves both GMV, ctr and IPV by 19.7\%, 8.1\% and 8.3\% respectively compared to Item-based CF. Even though the DNN-based LTR does not explicitly optimizing the GMV, it has 19.7\% improvements in GMV compared to item-based CF. By explicitly mapping the value of different actions into the GMV, our value-based RL method can bring another 6.8\% improvement on GMV compared to DNN-based LTR. The improvements on ctr and IPV is 0.3\% and 0.4\% respectively. Online test is done in real-world system (involving 2 hundred million clicks per day, 7 days). Based on significance test (p<0.005), the improvement are all significant and have business values for E-commercial platform, because every thousandth of improvement on ctr means hundreds of millions of extra clicks in real-world systems.

\begin{table}[t]
\vspace{5pt}
\caption{Offline performance when considering adding to cart and adding to wishlist (p-value < 0.005).}
\centering
\begin{adjustbox}{width=\linewidth}
\begin{tabular}{lrrrr}
\toprule
 &  E[GMV]  & \tabincell{c}{Precision\\@20(\%)} & \tabincell{c}{Recall\\@20(\%)} & \tabincell{c}{MAP\\(\%)}\\ 
 \midrule
\tabincell{l}{click}  & 42.2 &  3.67 & 34.5& 12.1  \\ 
\tabincell{l}{click,cart,fav}  & \tabincell{r}{43.5\\(3.1\%)} & \tabincell{r}{3.88\\(5.7\%) } &  \tabincell{r}{35.8\\(3.8\%)} & \tabincell{r}{12.9\\(6.6\%)} \\
\bottomrule
\end{tabular}
\end{adjustbox}
\label{tab:valuecartfav}
\end{table}

\subsection{Click vs. Add to Cart and Add to Wishlist}
In this part, we take adding to cart and adding to wishlist into consideration to evaluate the performance of our value-based model. Results in Table~\ref{tab:valuecartfav} show offline performance when considering the value of adding to cart and adding to wishlist actions into the model. The improvement on expected GMV is 3.1\%. The improvements on precision, recall and MAP are 5.7\%, 3.8\% and 6.6\% respectively. It means adding to cart and adding to wishlist actions are useful for the value-based profit maximization.

\section{Conclusions and Future Work}
\label{sec:conclusion}
Great advances have been achieved by existing research to improve the accuracy of rating prediction and the quality of top-k recommendation lists. However, the economic value of recommendation is rarely studied. For large-scale commercial recommendation system, state-of-the-art algorithms seldom maximize the final revenue/profit of the system. To eliminate this gap, we propose value-aware recommendation to maximize the profit of commercial recommendation system directly. Specifically, we generalize the basic concept of click conversion rate (CVR) to the conversation rate of an arbitrary user action on the platform (XVR), where different actions of users(\textit{click, add to cart, add to wishlist, purchase})can be monetized into the profit of the system. Then we use reinforcement learning (RL) to maximize the profit whose reward is the aggregated monetized user actions. Both offline and online experiments show that our value-based RL model not only performs better on traditional metrics such as precision, recall and MAP, but also greatly improves the final profit than existing methods. This paper acts as the first step towards value-aware recommendation and further improvement can be achieved by designing more powerful features and RL models in the future.

\vspace{2pt}



\begin{thebibliography}{00}


\ifx \showCODEN    \undefined \def \showCODEN     #1{\unskip}     \fi
\ifx \showDOI      \undefined \def \showDOI       #1{#1}\fi
\ifx \showISBNx    \undefined \def \showISBNx     #1{\unskip}     \fi
\ifx \showISBNxiii \undefined \def \showISBNxiii  #1{\unskip}     \fi
\ifx \showISSN     \undefined \def \showISSN      #1{\unskip}     \fi
\ifx \showLCCN     \undefined \def \showLCCN      #1{\unskip}     \fi
\ifx \shownote     \undefined \def \shownote      #1{#1}          \fi
\ifx \showarticletitle \undefined \def \showarticletitle #1{#1}   \fi
\ifx \showURL      \undefined \def \showURL       {\relax}        \fi
\providecommand\bibfield[2]{#2}
\providecommand\bibinfo[2]{#2}
\providecommand\natexlab[1]{#1}
\providecommand\showeprint[2][]{arXiv:#2}

\bibitem[\protect\citeauthoryear{??}{vam}{2017}]%
        {vams}
 \bibinfo{year}{2017}\natexlab{}.
\newblock \bibinfo{title}{VAMS2017}.
\newblock   (\bibinfo{year}{2017}).
\newblock
\showURL{%
\url{https://vams2017.wordpress.com/}}


\bibitem[\protect\citeauthoryear{Agarwal, Hosanagar, and Smith}{Agarwal
  et~al\mbox{.}}{2011}]%
        {agarwal2011location}
\bibfield{author}{\bibinfo{person}{Ashish Agarwal}, \bibinfo{person}{Kartik
  Hosanagar}, {and} \bibinfo{person}{Michael~D Smith}.}
  \bibinfo{year}{2011}\natexlab{}.
\newblock \showarticletitle{Location, location, location: An analysis of
  profitability of position in online advertising markets}.
\newblock \bibinfo{journal}{{\em Journal of marketing research\/}}
  \bibinfo{volume}{48}, \bibinfo{number}{6} (\bibinfo{year}{2011}),
  \bibinfo{pages}{1057--1073}.
\newblock


\bibitem[\protect\citeauthoryear{Broder, Fontoura, Josifovski, and
  Riedel}{Broder et~al\mbox{.}}{2007}]%
        {broder2007semantic}
\bibfield{author}{\bibinfo{person}{Andrei Broder}, \bibinfo{person}{Marcus
  Fontoura}, \bibinfo{person}{Vanja Josifovski}, {and} \bibinfo{person}{Lance
  Riedel}.} \bibinfo{year}{2007}\natexlab{}.
\newblock \showarticletitle{A semantic approach to contextual advertising}. In
  \bibinfo{booktitle}{{\em Proceedings of the 30th annual international ACM
  SIGIR conference on Research and development in information retrieval}}. ACM,
  \bibinfo{pages}{559--566}.
\newblock


\bibitem[\protect\citeauthoryear{Chen, Yu, Da, Tan, Huang, and Tang}{Chen
  et~al\mbox{.}}{2018}]%
        {kddali}
\bibfield{author}{\bibinfo{person}{Shi-Yong Chen}, \bibinfo{person}{Yang Yu},
  \bibinfo{person}{Qing Da}, \bibinfo{person}{Jun Tan},
  \bibinfo{person}{Hai-Kuan Huang}, {and} \bibinfo{person}{Hai-Hong Tang}.}
  \bibinfo{year}{2018}\natexlab{}.
\newblock \showarticletitle{Stabilizing reinforcement learning in dynamic
  environment with application to online recommendation}. In
  \bibinfo{booktitle}{{\em Proceedings of the 24th ACM SIGKDD International
  Conference on Knowledge Discovery \& Data Mining}}. ACM,
  \bibinfo{pages}{1187--1196}.
\newblock


\bibitem[\protect\citeauthoryear{Cremonesi, Koren, and Turrin}{Cremonesi
  et~al\mbox{.}}{2010}]%
        {cremonesi2010performance}
\bibfield{author}{\bibinfo{person}{Paolo Cremonesi}, \bibinfo{person}{Yehuda
  Koren}, {and} \bibinfo{person}{Roberto Turrin}.}
  \bibinfo{year}{2010}\natexlab{}.
\newblock \showarticletitle{Performance of recommender algorithms on top-n
  recommendation tasks}. In \bibinfo{booktitle}{{\em Proceedings of the fourth
  ACM conference on Recommender systems}}. ACM, \bibinfo{pages}{39--46}.
\newblock


\bibitem[\protect\citeauthoryear{Dave, Varma, et~al\mbox{.}}{Dave
  et~al\mbox{.}}{2014}]%
        {dave2014computational}
\bibfield{author}{\bibinfo{person}{Kushal Dave}, \bibinfo{person}{Vasudeva
  Varma}, {et~al\mbox{.}}} \bibinfo{year}{2014}\natexlab{}.
\newblock \showarticletitle{Computational advertising: Techniques for targeting
  relevant ads}.
\newblock \bibinfo{journal}{{\em Foundations and Trends{\textregistered} in
  Information Retrieval\/}} \bibinfo{volume}{8}, \bibinfo{number}{4--5}
  (\bibinfo{year}{2014}), \bibinfo{pages}{263--418}.
\newblock


\bibitem[\protect\citeauthoryear{Dietmar~Jannach}{Dietmar~Jannach}{2017}]%
        {value-aware1}
\bibfield{author}{\bibinfo{person}{Gediminas~Adomavicius Dietmar~Jannach}.}
  \bibinfo{year}{2017}\natexlab{}.
\newblock \showarticletitle{Price and Profit Awareness in Recommender Systems}.
\newblock \bibinfo{journal}{{\em arXiv preprint arXiv:1801.00209\/}}
  (\bibinfo{year}{2017}).
\newblock


\bibitem[\protect\citeauthoryear{Ekstrand, Riedl, Konstan,
  et~al\mbox{.}}{Ekstrand et~al\mbox{.}}{2011}]%
        {ekstrand2011collaborative}
\bibfield{author}{\bibinfo{person}{Michael~D Ekstrand}, \bibinfo{person}{John~T
  Riedl}, \bibinfo{person}{Joseph~A Konstan}, {et~al\mbox{.}}}
  \bibinfo{year}{2011}\natexlab{}.
\newblock \showarticletitle{Collaborative filtering recommender systems}.
\newblock \bibinfo{journal}{{\em Foundations and Trends{\textregistered} in
  Human--Computer Interaction\/}} \bibinfo{volume}{4}, \bibinfo{number}{2}
  (\bibinfo{year}{2011}), \bibinfo{pages}{81--173}.
\newblock


\bibitem[\protect\citeauthoryear{Feng, Li, Huang, Liu, Ou, Wang, and Zhu}{Feng
  et~al\mbox{.}}{2018}]%
        {feng2018learning}
\bibfield{author}{\bibinfo{person}{Jun Feng}, \bibinfo{person}{Heng Li},
  \bibinfo{person}{Minlie Huang}, \bibinfo{person}{Shichen Liu},
  \bibinfo{person}{Wenwu Ou}, \bibinfo{person}{Zhirong Wang}, {and}
  \bibinfo{person}{Xiaoyan Zhu}.} \bibinfo{year}{2018}\natexlab{}.
\newblock \showarticletitle{Learning to Collaborate: Multi-Scenario Ranking via
  Multi-Agent Reinforcement Learning}. In \bibinfo{booktitle}{{\em Proceedings
  of the 2018 World Wide Web Conference on World Wide Web}}. International
  World Wide Web Conferences Steering Committee, \bibinfo{pages}{1939--1948}.
\newblock


\bibitem[\protect\citeauthoryear{Ghose and Yang}{Ghose and Yang}{2009}]%
        {ghose2009empirical}
\bibfield{author}{\bibinfo{person}{Anindya Ghose} {and} \bibinfo{person}{Sha
  Yang}.} \bibinfo{year}{2009}\natexlab{}.
\newblock \showarticletitle{An empirical analysis of search engine advertising:
  Sponsored search in electronic markets}.
\newblock \bibinfo{journal}{{\em Management science\/}} \bibinfo{volume}{55},
  \bibinfo{number}{10} (\bibinfo{year}{2009}), \bibinfo{pages}{1605--1622}.
\newblock


\bibitem[\protect\citeauthoryear{Goldfarb and Tucker}{Goldfarb and
  Tucker}{2011}]%
        {goldfarb2011online}
\bibfield{author}{\bibinfo{person}{Avi Goldfarb} {and}
  \bibinfo{person}{Catherine Tucker}.} \bibinfo{year}{2011}\natexlab{}.
\newblock \showarticletitle{Online display advertising: Targeting and
  obtrusiveness}.
\newblock \bibinfo{journal}{{\em Marketing Science\/}} \bibinfo{volume}{30},
  \bibinfo{number}{3} (\bibinfo{year}{2011}), \bibinfo{pages}{389--404}.
\newblock


\bibitem[\protect\citeauthoryear{Gunawardana and Shani}{Gunawardana and
  Shani}{2009}]%
        {gunawardana2009survey}
\bibfield{author}{\bibinfo{person}{Asela Gunawardana} {and}
  \bibinfo{person}{Guy Shani}.} \bibinfo{year}{2009}\natexlab{}.
\newblock \showarticletitle{A survey of accuracy evaluation metrics of
  recommendation tasks}.
\newblock \bibinfo{journal}{{\em Journal of Machine Learning Research\/}}
  \bibinfo{volume}{10}, \bibinfo{number}{Dec} (\bibinfo{year}{2009}),
  \bibinfo{pages}{2935--2962}.
\newblock


\bibitem[\protect\citeauthoryear{He, Liao, Zhang, Nie, Hu, and Chua}{He
  et~al\mbox{.}}{2017}]%
        {he2017neural}
\bibfield{author}{\bibinfo{person}{Xiangnan He}, \bibinfo{person}{Lizi Liao},
  \bibinfo{person}{Hanwang Zhang}, \bibinfo{person}{Liqiang Nie},
  \bibinfo{person}{Xia Hu}, {and} \bibinfo{person}{Tat-Seng Chua}.}
  \bibinfo{year}{2017}\natexlab{}.
\newblock \showarticletitle{Neural collaborative filtering}. In
  \bibinfo{booktitle}{{\em Proceedings of the 26th International Conference on
  World Wide Web}}. International World Wide Web Conferences Steering
  Committee, \bibinfo{pages}{173--182}.
\newblock


\bibitem[\protect\citeauthoryear{Herlocker, Konstan, Borchers, and
  Riedl}{Herlocker et~al\mbox{.}}{2017}]%
        {herlocker2017algorithmic}
\bibfield{author}{\bibinfo{person}{Jonathan~L Herlocker},
  \bibinfo{person}{Joseph~A Konstan}, \bibinfo{person}{Al Borchers}, {and}
  \bibinfo{person}{John Riedl}.} \bibinfo{year}{2017}\natexlab{}.
\newblock \showarticletitle{An algorithmic framework for performing
  collaborative filtering}. In \bibinfo{booktitle}{{\em ACM SIGIR Forum}},
  Vol.~\bibinfo{volume}{51}. ACM, \bibinfo{pages}{227--234}.
\newblock


\bibitem[\protect\citeauthoryear{Hu, Da, Zeng, Yu, and Xu}{Hu
  et~al\mbox{.}}{2018}]%
        {ecommerce}
\bibfield{author}{\bibinfo{person}{Yujing Hu}, \bibinfo{person}{Qing Da},
  \bibinfo{person}{Anxiang Zeng}, \bibinfo{person}{Yang Yu}, {and}
  \bibinfo{person}{Yinghui Xu}.} \bibinfo{year}{2018}\natexlab{}.
\newblock \showarticletitle{Reinforcement Learning to Rank in E-Commerce Search
  Engine: Formalization, Analysis, and Application}. In
  \bibinfo{booktitle}{{\em Proceedings of the 24th ACM SIGKDD International
  Conference on Knowledge Discovery \& Data Mining}}. ACM.
\newblock


\bibitem[\protect\citeauthoryear{Koren, Bell, and Volinsky}{Koren
  et~al\mbox{.}}{2009}]%
        {koren2009matrix}
\bibfield{author}{\bibinfo{person}{Yehuda Koren}, \bibinfo{person}{Robert
  Bell}, {and} \bibinfo{person}{Chris Volinsky}.}
  \bibinfo{year}{2009}\natexlab{}.
\newblock \showarticletitle{Matrix factorization techniques for recommender
  systems}.
\newblock \bibinfo{journal}{{\em Computer\/}} \bibinfo{number}{8}
  (\bibinfo{year}{2009}), \bibinfo{pages}{30--37}.
\newblock


\bibitem[\protect\citeauthoryear{Lee and Seung}{Lee and Seung}{2001}]%
        {lee2001algorithms}
\bibfield{author}{\bibinfo{person}{Daniel~D Lee} {and}
  \bibinfo{person}{H~Sebastian Seung}.} \bibinfo{year}{2001}\natexlab{}.
\newblock \showarticletitle{Algorithms for non-negative matrix factorization}.
  In \bibinfo{booktitle}{{\em Advances in neural information processing
  systems}}. \bibinfo{pages}{556--562}.
\newblock


\bibitem[\protect\citeauthoryear{Lee, Orten, Dasdan, and Li}{Lee
  et~al\mbox{.}}{2012}]%
        {lee2012estimating}
\bibfield{author}{\bibinfo{person}{Kuang-chih Lee}, \bibinfo{person}{Burkay
  Orten}, \bibinfo{person}{Ali Dasdan}, {and} \bibinfo{person}{Wentong Li}.}
  \bibinfo{year}{2012}\natexlab{}.
\newblock \showarticletitle{Estimating conversion rate in display advertising
  from past erformance data}. In \bibinfo{booktitle}{{\em Proceedings of the
  18th ACM SIGKDD international conference on Knowledge discovery and data
  mining}}. ACM, \bibinfo{pages}{768--776}.
\newblock


\bibitem[\protect\citeauthoryear{Liu}{Liu}{2009}]%
        {Liu:2009:LRI:1618303.1618304}
\bibfield{author}{\bibinfo{person}{Tie-Yan Liu}.}
  \bibinfo{year}{2009}\natexlab{}.
\newblock \showarticletitle{Learning to Rank for Information Retrieval}.
\newblock \bibinfo{journal}{{\em Found. Trends Inf. Retr.\/}}
  \bibinfo{volume}{3}, \bibinfo{number}{3} (\bibinfo{date}{March}
  \bibinfo{year}{2009}), \bibinfo{pages}{225--331}.
\newblock
\showISSN{1554-0669}


\bibitem[\protect\citeauthoryear{Ma, Zhao, Huang, Wang, Hu, Zhu, and Gai}{Ma
  et~al\mbox{.}}{2018}]%
        {cvr-ali}
\bibfield{author}{\bibinfo{person}{Xiao Ma}, \bibinfo{person}{Liqin Zhao},
  \bibinfo{person}{Guan Huang}, \bibinfo{person}{Zhi Wang},
  \bibinfo{person}{Zelin Hu}, \bibinfo{person}{Xiaoqiang Zhu}, {and}
  \bibinfo{person}{Kun Gai}.} \bibinfo{year}{2018}\natexlab{}.
\newblock \showarticletitle{Entire Space Multi-Task Model: An Effective
  Approach for Estimating Post-Click Conversion Rate}. In
  \bibinfo{booktitle}{{\em Proceedings of the 41st International ACM SIGIR
  Conference on Research \& Development in Information Retrieval}}. ACM,
  \bibinfo{pages}{1137--1140}.
\newblock


\bibitem[\protect\citeauthoryear{Mnih and Salakhutdinov}{Mnih and
  Salakhutdinov}{2008}]%
        {mnih2008probabilistic}
\bibfield{author}{\bibinfo{person}{Andriy Mnih} {and} \bibinfo{person}{Ruslan~R
  Salakhutdinov}.} \bibinfo{year}{2008}\natexlab{}.
\newblock \showarticletitle{Probabilistic matrix factorization}. In
  \bibinfo{booktitle}{{\em Advances in neural information processing systems}}.
\newblock


\bibitem[\protect\citeauthoryear{Pazzani and Billsus}{Pazzani and
  Billsus}{2007}]%
        {pazzani2007content}
\bibfield{author}{\bibinfo{person}{Michael~J Pazzani} {and}
  \bibinfo{person}{Daniel Billsus}.} \bibinfo{year}{2007}\natexlab{}.
\newblock \showarticletitle{Content-based recommendation systems}.
\newblock In \bibinfo{booktitle}{{\em The adaptive web}}.
  \bibinfo{publisher}{Springer}, \bibinfo{pages}{325--341}.
\newblock


\bibitem[\protect\citeauthoryear{Rendle and Freudenthaler}{Rendle and
  Freudenthaler}{2014}]%
        {rendle2014improving}
\bibfield{author}{\bibinfo{person}{Steffen Rendle} {and}
  \bibinfo{person}{Christoph Freudenthaler}.} \bibinfo{year}{2014}\natexlab{}.
\newblock \showarticletitle{Improving pairwise learning for item recommendation
  from implicit feedback}. In \bibinfo{booktitle}{{\em Proceedings of the 7th
  ACM international conference on Web search and data mining}}. ACM,
  \bibinfo{pages}{273--282}.
\newblock


\bibitem[\protect\citeauthoryear{Salimans, Ho, Chen, Sidor, and
  Sutskever}{Salimans et~al\mbox{.}}{2017}]%
        {es}
\bibfield{author}{\bibinfo{person}{Tim Salimans}, \bibinfo{person}{Jonathan
  Ho}, \bibinfo{person}{Xi Chen}, \bibinfo{person}{Szymon Sidor}, {and}
  \bibinfo{person}{Ilya Sutskever}.} \bibinfo{year}{2017}\natexlab{}.
\newblock \showarticletitle{Evolution strategies as a scalable alternative to
  reinforcement learning}.
\newblock \bibinfo{journal}{{\em arXiv preprint arXiv:1703.03864\/}}
  (\bibinfo{year}{2017}).
\newblock


\bibitem[\protect\citeauthoryear{Sarwar, Karypis, Konstan, and Riedl}{Sarwar
  et~al\mbox{.}}{2001}]%
        {sarwar2001item}
\bibfield{author}{\bibinfo{person}{Badrul Sarwar}, \bibinfo{person}{George
  Karypis}, \bibinfo{person}{Joseph Konstan}, {and} \bibinfo{person}{John
  Riedl}.} \bibinfo{year}{2001}\natexlab{}.
\newblock \showarticletitle{Item-based collaborative filtering recommendation
  algorithms}. In \bibinfo{booktitle}{{\em Proceedings of the 10th
  international conference on World Wide Web}}. ACM, \bibinfo{pages}{285--295}.
\newblock


\bibitem[\protect\citeauthoryear{Shani, Heckerman, and Brafman}{Shani
  et~al\mbox{.}}{2005}]%
        {yxr25}
\bibfield{author}{\bibinfo{person}{Guy Shani}, \bibinfo{person}{David
  Heckerman}, {and} \bibinfo{person}{Ronen~I Brafman}.}
  \bibinfo{year}{2005}\natexlab{}.
\newblock \showarticletitle{An MDP-based recommender system}.
\newblock \bibinfo{journal}{{\em Journal of Machine Learning Research\/}}
  \bibinfo{volume}{6}, \bibinfo{number}{Sep} (\bibinfo{year}{2005}),
  \bibinfo{pages}{1265--1295}.
\newblock


\bibitem[\protect\citeauthoryear{SIGKDD}{SIGKDD}{2015}]%
        {jiang2015life}
\bibfield{author}{\bibinfo{person}{SIGKDD}.} \bibinfo{year}{2015}\natexlab{}.
\newblock \showarticletitle{Life-stage Prediction for Product Recommendation in
  E-commerce}. ACM, \bibinfo{pages}{1879--1888}.
\newblock


\bibitem[\protect\citeauthoryear{Sutton, Barto, et~al\mbox{.}}{Sutton
  et~al\mbox{.}}{1998}]%
        {RL}
\bibfield{author}{\bibinfo{person}{Richard~S Sutton}, \bibinfo{person}{Andrew~G
  Barto}, {et~al\mbox{.}}} \bibinfo{year}{1998}\natexlab{}.
\newblock \bibinfo{booktitle}{{\em Reinforcement learning: An introduction}}.
\newblock \bibinfo{publisher}{MIT press}.
\newblock


\bibitem[\protect\citeauthoryear{Taghipour and Kardan}{Taghipour and
  Kardan}{2008}]%
        {yxr27}
\bibfield{author}{\bibinfo{person}{Nima Taghipour} {and} \bibinfo{person}{Ahmad
  Kardan}.} \bibinfo{year}{2008}\natexlab{}.
\newblock \showarticletitle{A hybrid web recommender system based on
  q-learning}. In \bibinfo{booktitle}{{\em Proceedings of the 2008 ACM
  symposium on Applied computing}}. ACM, \bibinfo{pages}{1164--1168}.
\newblock


\bibitem[\protect\citeauthoryear{Wang, Wang, and Yeung}{Wang
  et~al\mbox{.}}{2015}]%
        {wang2015collaborative}
\bibfield{author}{\bibinfo{person}{Hao Wang}, \bibinfo{person}{Naiyan Wang},
  {and} \bibinfo{person}{Dit-Yan Yeung}.} \bibinfo{year}{2015}\natexlab{}.
\newblock \showarticletitle{Collaborative deep learning for recommender
  systems}. In \bibinfo{booktitle}{{\em Proceedings of the 21th ACM SIGKDD
  International Conference on Knowledge Discovery and Data Mining}}. ACM,
  \bibinfo{pages}{1235--1244}.
\newblock


\bibitem[\protect\citeauthoryear{Xia, Jiang, Sun, Zhang, Wang, and Sui}{Xia
  et~al\mbox{.}}{2018}]%
        {xia2018modeling}
\bibfield{author}{\bibinfo{person}{Qiaolin Xia}, \bibinfo{person}{Peng Jiang},
  \bibinfo{person}{Fei Sun}, \bibinfo{person}{Yi Zhang},
  \bibinfo{person}{Xiaobo Wang}, {and} \bibinfo{person}{Zhifang Sui}.}
  \bibinfo{year}{2018}\natexlab{}.
\newblock \showarticletitle{Modeling Consumer Buying Decision for
  Recommendation Based on Multi-Task Deep Learning}. In
  \bibinfo{booktitle}{{\em CIKM}}. ACM, \bibinfo{pages}{1703--1706}.
\newblock


\bibitem[\protect\citeauthoryear{Zhang, Yao, and Sun}{Zhang
  et~al\mbox{.}}{2017}]%
        {zhang2017deep}
\bibfield{author}{\bibinfo{person}{Shuai Zhang}, \bibinfo{person}{Lina Yao},
  {and} \bibinfo{person}{Aixin Sun}.} \bibinfo{year}{2017}\natexlab{}.
\newblock \showarticletitle{Deep learning based recommender system: A survey
  and new perspectives}.
\newblock \bibinfo{journal}{{\em arXiv preprint arXiv:1707.07435\/}}
  (\bibinfo{year}{2017}).
\newblock


\bibitem[\protect\citeauthoryear{Zhang, Zhao, Zhang, Friedman, Zhang, Liu, and
  Ma}{Zhang et~al\mbox{.}}{2016}]%
        {zhang2016economic}
\bibfield{author}{\bibinfo{person}{Yongfeng Zhang}, \bibinfo{person}{Qi Zhao},
  \bibinfo{person}{Yi Zhang}, \bibinfo{person}{Daniel Friedman},
  \bibinfo{person}{Min Zhang}, \bibinfo{person}{Yiqun Liu}, {and}
  \bibinfo{person}{Shaoping Ma}.} \bibinfo{year}{2016}\natexlab{}.
\newblock \showarticletitle{Economic recommendation with surplus maximization}.
  In \bibinfo{booktitle}{{\em WWW}}.
\newblock


\bibitem[\protect\citeauthoryear{Zhao, Zhang, Zhang, and Friedman}{Zhao
  et~al\mbox{.}}{2017b}]%
        {zhao2017multi}
\bibfield{author}{\bibinfo{person}{Qi Zhao}, \bibinfo{person}{Yongfeng Zhang},
  \bibinfo{person}{Yi Zhang}, {and} \bibinfo{person}{Daniel Friedman}.}
  \bibinfo{year}{2017}\natexlab{b}.
\newblock \showarticletitle{Multi-product utility maximization for economic
  recommendation}. In \bibinfo{booktitle}{{\em Proceedings of the Tenth ACM
  International Conference on Web Search and Data Mining}}. ACM,
  \bibinfo{pages}{435--443}.
\newblock


\bibitem[\protect\citeauthoryear{Zhao, Xia, Zhang, Ding, Yin, and Tang}{Zhao
  et~al\mbox{.}}{2018a}]%
        {pagewise}
\bibfield{author}{\bibinfo{person}{Xiangyu Zhao}, \bibinfo{person}{Long Xia},
  \bibinfo{person}{Liang Zhang}, \bibinfo{person}{Zhuoye Ding},
  \bibinfo{person}{Dawei Yin}, {and} \bibinfo{person}{Jiliang Tang}.}
  \bibinfo{year}{2018}\natexlab{a}.
\newblock \showarticletitle{Deep Reinforcement Learning for Page-wise
  Recommendations}. In \bibinfo{booktitle}{{\em Proceedings of the 12th ACM
  Conference on Recommender Systems}}. ACM.
\newblock


\bibitem[\protect\citeauthoryear{Zhao, Zhang, Ding, Xia, Tang, and Yin}{Zhao
  et~al\mbox{.}}{2018b}]%
        {negative}
\bibfield{author}{\bibinfo{person}{Xiangyu Zhao}, \bibinfo{person}{Liang
  Zhang}, \bibinfo{person}{Zhuoye Ding}, \bibinfo{person}{Long Xia},
  \bibinfo{person}{Jiliang Tang}, {and} \bibinfo{person}{Dawei Yin}.}
  \bibinfo{year}{2018}\natexlab{b}.
\newblock \showarticletitle{{Recommendation with Negative Feedback via Pairwise
  Deep Reinforcement Learning}}.
\newblock \bibinfo{journal}{{\em Proceedings of the 24th ACM SIGKDD Conference
  on Knowledge Discovery and Data Mining\/}} (\bibinfo{year}{2018}).
\newblock


\bibitem[\protect\citeauthoryear{Zhao, Zhang, Ding, Yin, Zhao, and Tang}{Zhao
  et~al\mbox{.}}{2017a}]%
        {listwise}
\bibfield{author}{\bibinfo{person}{Xiangyu Zhao}, \bibinfo{person}{Liang
  Zhang}, \bibinfo{person}{Zhuoye Ding}, \bibinfo{person}{Dawei Yin},
  \bibinfo{person}{Yihong Zhao}, {and} \bibinfo{person}{Jiliang Tang}.}
  \bibinfo{year}{2017}\natexlab{a}.
\newblock \showarticletitle{Deep Reinforcement Learning for List-wise
  Recommendations}.
\newblock \bibinfo{journal}{{\em arXiv preprint arXiv:1801.00209\/}}
  (\bibinfo{year}{2017}).
\newblock


\bibitem[\protect\citeauthoryear{Zheng}{Zheng}{2017}]%
        {value-aware2}
\bibfield{author}{\bibinfo{person}{Yong Zheng}.}
  \bibinfo{year}{2017}\natexlab{}.
\newblock \showarticletitle{Multi-Stakeholder Recommendation: Applications and
  Challenges}.
\newblock \bibinfo{journal}{{\em arXiv preprint arXiv:1707.08913\/}}
  (\bibinfo{year}{2017}).
\newblock


\bibitem[\protect\citeauthoryear{Zhu, Jin, Tan, Pan, Zeng, Li, and Gai}{Zhu
  et~al\mbox{.}}{}]%
        {cvr-ad}
\bibfield{author}{\bibinfo{person}{Han Zhu}, \bibinfo{person}{Junqi Jin},
  \bibinfo{person}{Chang Tan}, \bibinfo{person}{Fei Pan},
  \bibinfo{person}{Yifan Zeng}, \bibinfo{person}{Han Li}, {and}
  \bibinfo{person}{Kun Gai}.}
\newblock \showarticletitle{Optimized cost per click in taobao display
  advertising}.
\newblock


\end{thebibliography}

\end{document}